\documentstyle         {mn}

\newif\ifAMStwofonts

\ifoldfss

  \ifCUPmtlplainloaded \else
    \NewTextAlphabet{textbfit} {cmbxti10} {}
    \NewTextAlphabet{textbfss} {cmssbx10} {}
    \NewMathAlphabet{mathbfit} {cmbxti10} {} 
    \NewMathAlphabet{mathbfss} {cmssbx10} {} 
  \fi
  \ifAMStwofonts
    \ifCUPmtlplainloaded \else
      \NewSymbolFont{upmath} {eurm10}
      \NewSymbolFont{AMSa} {msam10}
      \NewMathSymbol{\upi}     {0}{upmath}{19}
      \NewMathSymbol{\umu}     {0}{upmath}{16}
      \NewMathSymbol{\upartial}{0}{upmath}{40}
      \NewMathSymbol{\leqslant}{3}{AMSa}{36}
      \NewMathSymbol{\geqslant}{3}{AMSa}{3E}

       \let\le=\leqslant
       
    \fi
  \fi
\fi 

\ifnfssone
  \newmathalphabet{\mathit}
  \addtoversion{normal}{\mathit}{cmr}{m}{it}
  \addtoversion{bold}{\mathit}{cmr}{bx}{it}
  \newmathalphabet{\mathbfit} 
  \addtoversion{normal}{\mathbfit}{cmr}{bx}{it}
  \addtoversion{bold}{\mathbfit}{cmr}{bx}{it}
  \newmathalphabet{\mathbfss} 
  \addtoversion{normal}{\mathbfss}{cmss}{bx}{n}
  \addtoversion{bold}{\mathbfss}{cmss}{bx}{n}
  \ifAMStwofonts
    \ifCUPmtlplainloaded \else
      \UseAMStwoboldmath
      \makeatletter
      \new@mathgroup\upmath@group
      \define@mathgroup\mv@normal\upmath@group{eur}{m}{n}
      \define@mathgroup\mv@bold\upmath@group{eur}{b}{n}
      \edef\UPM{\hexnumber\upmath@group}
      \new@mathgroup\amsa@group
      \define@mathgroup\mv@normal\amsa@group{msa}{m}{n}
      \define@mathgroup\mv@bold\amsa@group{msa}{m}{n}
      \edef\AMSa{\hexnumber\amsa@group}
      \makeatother
      \mathchardef\upi="0\UPM19
      \mathchardef\umu="0\UPM16
      \mathchardef\upartial="0\UPM40
      \mathchardef\leqslant="3\AMSa36
      \mathchardef\geqslant="3\AMSa3E

       \let\le=\leqslant

    \fi
  \fi
\fi 

\ifnfsstwo
  \DeclareMathAlphabet{\mathbfit}{OT1}{cmr}{bx}{it}
  \SetMathAlphabet\mathbfit{bold}{OT1}{cmr}{bx}{it}
  \DeclareMathAlphabet{\mathbfss}{OT1}{cmss}{bx}{n}
  \SetMathAlphabet\mathbfss{bold}{OT1}{cmss}{bx}{n}
  \ifAMStwofonts
    \ifCUPmtlplainloaded \else
      \DeclareSymbolFont{UPM}{U}{eur}{m}{n}
      \SetSymbolFont{UPM}{bold}{U}{eur}{b}{n}
      \DeclareSymbolFont{AMSa}{U}{msa}{m}{n}
      \DeclareMathSymbol{\upi}{0}{UPM}{"19}
      \DeclareMathSymbol{\umu}{0}{UPM}{"16}
      \DeclareMathSymbol{\upartial}{0}{UPM}{"40}
      \DeclareMathSymbol{\leqslant}{3}{AMSa}{"36}
      \DeclareMathSymbol{\geqslant}{3}{AMSa}{"3E}

       \let\le=\leqslant

    \fi
  \fi
\fi 

\ifCUPmtlplainloaded \else
  \ifAMStwofonts \else 
    \def\upi{\pi}
    \def\umu{\mu}
    \def\upartial{\partial}
  \fi
\fi

\title [Discovering Galactic Planets by Microlensing]
{Discovering Galactic Planets by Gravitational Microlensing:
	Magnification Patterns and Light Curves}
\author[Joachim Wambsganss]
       {Joachim Wambsganss \\
	Astrophysikalisches Institut Potsdam, An der Sternwarte 16, 14482 Potsdam, Germany; e-mail: jwambsganss@aip.de }
\date{Accepted 1996 August 9.
      Received 1996 June 10}

\pagerange{\pageref{firstpage}--\pageref{lastpage}}
\pubyear{1996}

\begin{document}

\maketitle

\label{firstpage}

\begin{abstract}

The current searches for microlensing events towards the galactic 
bulge can be used to detect planetary companions around the 
lensing stars. 
The effect of such planets is a short-term modulation
on the smooth microlensing
lightcurve produced by the main lensing star.
Current and planned experiments should be sensitive enough 
to discover 
planets ranging  from Jupiter mass down to Earth mass.  
In order to be able to successfully  detect planets this way, 
it is necessary to accurately  and frequently  monitor a microlensing 
event photometrically, once it has been ``triggered". 

Here we present a large variety of two-dimensional magnification 
distributions for systems consisting of an ordinary  star and a 
planetary companion.  
We cover planet/star mass 
ratios from  $m_{pl}/M_* = 10^{-5}$ to $ 10^{-3}$. 
These limits
correspond roughly to $M_{Earth}$ and $M_{Jupiter}$, for a typical 
lens    mass of $M_* \approx 0.3 M_{\odot}$.
We explore a range of star-planet distances, with particular emphasis
on the case of ``resonant lensing", a situation in which the
planet is located at or very near the Einstein ring of the lensing star.

We show a wide selection of light curves 
-- one dimensional cuts through the magnification patterns -- to 
illustrate the broad range of possible light curve perturbations caused
by planets.  The strongest effects are to be expected for caustic 
crossings. But even tracks passing outside the caustics can have 
considerable effects on the light curves.  
The easiest detectable (projected)
distance range for the planets is between
about 0.6 and 1.6 Einstein radii. Planets in this distance range
produce caustics {\it inside} the Einstein ring of the star.
For a lensing star with a mass of about $0.3 M_{\odot}$ at 
a distance of 6 kpc and a source at 8 kpc,
this  corresponds to  physical distances between star and
planet of about   1 to 3 AU.

\end{abstract}


\section{Introduction}

The fraction of stars that have planetary companions
is unknown. It is certainly  not zero, and it can be as high as 100\%.
After the first discovery of a multiple planet system around pulsar 
PSR 1257+12 (Wolszczan \& Frail 1992; Wolszczan 1994),
recently various groups 
found ``planets" around nearby stars, e.g. around   
Gliese 229A, 47 Ursae Majoris, 70 Virginis and 51 Pegasi
(Mayor \& Queloz 1995; Nakajima et al. 1995; Butler \& Marcy 1996; 
Marcy \& Butler 1996).
These are reassuring observations, since it would have been bizarre
if the sun would be the only star with a planetary system, and 
we lived in such a ``special" place.

In the last few years, various groups have detected microlensing events
caused by foreground objects passing in front of stars in the
Large Magellanic Cloud or in the bulge of our Galaxy
(Alcock et al. 1993; Udalski et al. 1993; 
Alard et al. 1995;
Alard 1996).
The idea goes back to Paczy\'nski (1986) who suggested to use the 
gravitational lens effect of foreground objects on background stars 
for the search of dark matter in compact 
form in the Galactic Halo and Disk.
By now more than 100 such microlensing events are recorded 
with the bulk of the durations between 10 days and 100 days
(cf.  Paczy\'nski 1996a,b). 

Most of these microlensing events follow very nicely the
smooth symmetric light curves, as predicted by Paczy\'nski (1986)
or Griest (1991).
Meanwhile a couple of non-standard events have been observed as well. 
These are without any doubt   microlensed stars, 
but the lightcurve deviates clearly from
the ideal shape of an isolated lens acting on an isolated
background point source.
Examples of such ``perturbations"
are the effect of the parallax of the Earth due to its motion
around the sun during the event (Alcock et al. 1995),
or light curves caused by binary lenses (Udalski et al. 1994a;
Dominik \& Hirshfeld 1994; Mao \& Di Stefano 1995).
This proofs not only that the experiments are working in principle, 
but also that they are excellently doing the non-trivial job of 
accurately and frequently monitoring millions of stars and filtering
out the highly interesting fraction of about $10^{-7}$ light curves
affected by microlensing. And most importantly for our purpose here,
these experiments are able to find microlensed lightcurves that 
deviate from the simple lightcurve of an isolated point lens.

In this paper we will investigate the effects of
planetary companions to a star that acts as a microlens. 
Various aspects of microlensing by planets have 
been explored in the past. 
Mao and Paczy\'nski (1991)  were the first to suggest that these 
microlensing searches can be used to detect extra-solar planets.
Gould and Loeb (1992) quantitatively estimated the fraction 
of microlensed light curves that are affected by planets (of
Jupiter and Saturn mass), and the typical durations of the 
planetary perturbations.
Bolatto and Falco (1994) perform a more refined and
realistic analysis of the probability of detecting planetary
companions for mass ratios down to $10^{-3}$. 
Very recently  Bennett and Rhie (1996)  include
the effect of the finite size of the source and show and
emphasize that the microlensing searches
are able to detect planets down to the Earth mass. 
In a recent review, Paczy\'nski (1996b)  also investigates the odds for 
detecting earth mass planets with gravitational microlensing.

After rederiving some relevant numbers and orders of magnitude
in Chapter 2, 
we briefly present the ray shooting method in Chapter 3, 
which we use
to determine the magnification properties of a star-planet system.
A large variety of maps of the microlensing magnification as a 
function of the source location,  
produced by various planet/star mass ratios and a range of planet-star 
distances are shown in Chapter 4. 
Many example light curves -- one-dimensional cuts through these 
two-dimensional magnification distributions -- are presented to
illustrate the wide range of possible light curve perturbations.
In Chapter 5 we summarize our results and conclude with a brief
comparison between the various methods to discover planets.
%
%
%
%
%
\section{A Few Relevant Numbers}

\noindent

The natural length scale for microlensing is the Einstein ring radius

\begin{equation}
R_E = \sqrt{ { {4 G M }\over {c^2} } { {(D_S - D_L)D_L  } \over {D_S} } }
    = \sqrt{ { {4 G M }\over {c^2} } { D_S  (1 - d) \, d  } }
\end{equation}

\noindent
where $M$ is the mass of the lensing star, 
$D_L$ and $D_S$ are the distances to the lens and to the source,
respectively, $d = D_L/D_S$ is the distance ratio,  
and $G$ and $c$ are the gravitational constant and
the velocity of light.  
Parametrized by mass and distances,  the Einstein ring
radius in the lens plane can be expressed as follows:
\begin{equation}
	R_E      = 8.1 \ ( M / M_\odot )^{0.5}
                         ( D_S / 8 kpc   )^{0.5}
                         ( (1-d) \, d     )^{0.5}  
			 \ AU
\end{equation}
$$
	= 1.21 \times 10^{14}\ (M / M_\odot)^{0.5}
                         ( D_S / 8 kpc   )^{0.5}
                         ( (1-d) \, d     )^{0.5}  
			\ cm.
$$ 
In the source plane the physical size increases by the ratio
$D_S/D_L = d^{-1}$.
The Einstein angle, i.e. $R_E$ in  angular units, is given as

\begin{equation}
	\theta_E = R_E / D_L 
\end{equation}
$$
	= 1.0 \      (M / M_\odot)^{0.5}
                         ( D_L / 8 kpc   )^{-0.5}
                         ( 1-d      )^{0.5}  
			\  mas.
$$

The microlensing magnification  as a function of time is given as 
\begin{equation}
\mu(t) = { {u(t)^2 + 2} \over {u(t) \sqrt{u(t)^2 + 4}} },
\end{equation}
where $u(t)$ is the projected distance between lens and source
at time $t$ (in units of the Einstein radius).
For a typical transverse velocity of 
$v_\perp = 200 km/sec *  v_{\perp,200}$,
the time scale of a microlensing event produced by a star with
mass $M$ is given as
\begin{equation}
t_E = R_E / v_\perp 
\end{equation}
	$$
	= 69.9 \       (M / M_\odot)^{0.5}
		\      (  D_S / 8kpc)^{0.5}
		\      ( (1 - d) \, d  )^{0.5}
		\  \ \ v_{\perp,200}^{-1} 	
		\ \ \ days.
	$$
For  comparison with the duration of the ``perturbations" discussed
below, 
it is worthwhile to point out that a time
interval corresponding to 1\% of the  natural unit is
$$
t_{1\%} = 0.01 \times t_E \approx  7.3 (M / M_\odot)^{0.5} \ hours
$$
for the values of $v_\perp = 200 km/sec$ and 
$D_L = 6 kpc$, $D_S = 8 kpc$.

A stellar radius of $1 R_\odot$ 
transforms into an angular source size of
$\theta_S \approx 0.001 \, \theta_E$ for a
source distance of $ D_S = 8 kpc$ as used above. 
Such values show that it is 
justified to consider the finite extend of the sources
(cf. Witt \& Mao 1994; Witt 1995), 
rather than treating them as point sources.
In the examples of the lightcurves below, we use
angular source radii of $\theta_S  = 0.001 \, \theta_E$.

The mass of the Earth is about $3 \times 10^{-6} M_\odot$. Since
most potential lensing stars are somewhat less massive than the sun,
a mass ratio of $m_{pl} / M_* = 10^{-5}$ can be assumed
to roughly correspond to an Earth type planet.
Similarly, $m_{pl} / M_* = 10^{-4}$ and $10^{-3}$ are not too far from 
representing
the masses of Uranus ($M_{Uranus} = 4.4 \times 10^{-5} M_{\odot}$)
and Saturn ($M_{Saturn} = 2.9 \times 10^{-4} M_{\odot}$), relative 
to a low mass main sequence star. Jupiter's
mass would correspond to an even higher mass ratio
in that case
($M_{Jupiter} = 9.5 \times 10^{-4} M_{\odot}$) 

%
%
%
%
%
\section{The Method }

\noindent

In order to determine a two-dimensional magnification map
of the star-planet system in the source plane
we use the inverse ray-shooting technique 
(cf. Schneider \& Weiss 1986, 
Kayser, Refsdal \& Stabell 1986; 
Wambsganss 1990; Wambsganss et al. 1990). 
We ``shoot" light rays backwards from an observer 
through a lens plane that consists of a star with mass $M_*$ and a 
planetary  companion with mass 
$m_{pl}/M_* = 10^{-3}, 10^{-4}$ or $10^{-5}$.
The rays are collected in the source plane, and the density 
of the rays at a particular location  in the source plane
is directly proportional to the magnification at this point.

In order to obtain a lightcurve from such a two-dimensional pattern,
one has to consider a star with a certain brightness
profile serving as the ``source",
and -- as a first approximation --
assuming a straight track of the background star relative
to the lens-observer system. In other words, a light curve is a
one-dimensional cut through such a  magnification pattern, 
folded with a source profile of a background star.

Below, most of the magnification patterns and
light curves are displayed as the differences between  the case
``star plus planet" and ``star without planet". The advantage of this 
representation  is a gain in dynamical range. And -- after all --
this is the way those galactic planets
will be detected: as the differences
of an observed light curve and a theoretical one by an isolated lens
(Paczynski 1986, Griest 1991).

For the ray shooting, the planets are placed at various distances to the
main star. 
In Figure 1 the 
lens plane geometry 
of the various scenarios 
is illustrated.
The star is indicated with a `star' symbol at the origin of the
coordinate system, and the various  positions of the planets
investigated are marked with little crosses.  
The dashed line marks   the Einstein radius. 
The top panel (Figure 1a) illustrates the cases explored in detail in 
Section 4.1 and Figures 2, 3 and 4 
with eight planets at the following positions:
$x_{pl}/R_E \approx  0.57, 0.65, 0.74, 0.86, 1.16, 
1.34, 1.55,  1.77 $. These eight planets are 
mapped simultaneously for illustrative purposes.
The lens positions are  chosen such that the caustics appear
about equidistant in the source plane at 
$\theta/\theta_E = -1.2, -0.9, -0.6, -0.3, 0.3, 0.6, 0.9, 1.2$.
The middle panel (Fig. 1b) shows four planets at  positions 
$x_{pl}/R_E = 0.57, 0.86,  1.16 $ and 1.77.
They are investigated individually and
in detail in section 4.2, and displayed in Figures 5, 6 and  7. 
The lowest panel (Fig. 1c) with planet positions very close to the
Einstein ring radius illustrates the cases of ``resonant lensing"
looked at more closely in Figures 8, 9 and 10, and commented
on in section 4.3. (``Resonant lensing" is equivalent to an isolated 
lens with an external shear equal to unity, which is 
not treatable analytically.)

%
%
%
%
\section{Results: Magnification Patterns and Lightcurves}

Below we will present the magnification distributions and light
curves in three ways. The first is a combined  view of eight
planetary companions on the same scale as the stellar lensing, 
in order to see the relative importance and the qualitative 
properties (Section 4.1). 
In the second part  the lensing
effects for four different planet-star distances are explored 
individually in detail (Section 4.2), 
and in the last part, we investigate the case of ``resonant 
lensing", with four planet positions very close to the Einstein ring
(Section 4.3).

\subsection*{4.1 Qualitative Features: Eight Aligned Planets}

In Figure 2 the lensing effects  of planets at various distances to the
main lensing star are shown on the scale of the star. 
The magnification pattern in the source plane indicates
the magnification as a function of position: the darker the
gray, the higher the magnification. 
The black circle marks the Einstein ring of the star.  
There are small perturbations visible that are caused by individual
planets. For simplicity, we superposed the action 
of eight aligned planets at distances between 
$x_{pl} = 0.57 R_E$ and $x_{pl} = 1.77 R_E$,
mapped simultaneously for qualitative purposes.  
The geometric arrangement of the planets can be seen in Figure 1a. 
These lens positions 
 translate into a range of locations of the caustics in the 
 source plane between -1.2 and +1.2 in normalized units.
Figure 2a (left panel)
corresponds to planet masses of $m_{pl}/M_* = 10^{-3}$, 
Figure 2b (right panel) to  $m_{pl}/M_* = 10^{-4}$.

Most obvious is the changing 
appearance of the caustics depending on the
lens-star distance: 
For very large values of $x_{pl}$ (the rightmost perturbation
in the magnification pattern) the planet
caustic  -- an astroid with four cusps connected by four folds --
is very small and far outside the Einstein radius. It is obvious that
the planet causes additional magnification, the region at and around
the caustic  is darker than the environment.
For $x_{pl} \approx  1.618 R_E$ the caustic would be exactly on the 
Einstein
ring (not displayed here), and it keeps growing in size with decreasing
$x_{pl}$. 
For the planet with the second largest distance 
($x_{pl} \approx  1.55 R_E $) 
the caustics are already inside the Einstein
ring at $\theta = 0.9 \theta_E$; the next two planet positions
($x_{pl}/R_E \approx  1.34 $ and  1.16 ) correspond to caustic positions
$\theta = 0.6 \theta_E$ and $0.3 \theta_E$
For values of $x_{pl}$ approaching  $R_E$, the planet
caustic merges with the caustic of the main star at the center
(cf. section 4.3). 
For  planet positions $x_{pl}/R_E < 1.0$, the caustics lie on the
opposite side of the star, and 
the character of the caustics has changed: Now there are two
triangular caustics with an area of ``depression" in between. 
That means that for certain source positions the effect of an additional
planet around a microlensing star can be a ``de-magnification", compared
to the case of the star without the planet.
With still smaller distance planet-star, the two triangular
caustics get smaller, their relative separation gets larger.
The distance values are $x_{pl}/R_E \approx 0.86 $, 0.74,  0.65
for the caustic locations $\theta = -0.3 \theta_E$, 
                          $         -0.6 \theta_E$, 
                          $         -0.9 \theta_E$. 
At $x_{pl} \approx  0.618 R_E$ the pair of caustics would be at the 
distance of the Einstein radius ``on the other side" of the lens. 
The leftmost caustics displayed here at $\theta = -1.2 \theta_E$ 
correspond to $x_{pl} \approx 0.57 R_E$.
For still smaller planet-star distances, approaching 
$x_{pl} \approx 0$, the locations of the caustics approach infinity.
This  shows that the star-planet distance range which is most likely
to be detected as a perturbation of a microlensed lightcurve by
a star 
is roughly $0.5 R_E \le x_{pl} \le   2.0 R_E$.
This is the distance range for which the planet-star caustics are
located inside or near the Einstein ring radius of the star. 
\footnote{For planet distances much smaller than 0.5 $R_E$ or 
much larger than 2.0 $R_E$, the lensing signature 
is unlikely to be detected as a perturbation of the lightcurve 
of a lensed star, but rather as an individual microlensing event,
which will be more difficult in practice.}
With typical distances of $D_L = 6 kpc$ and
$D_S = 8 kpc$, this translates
into a physical range of (projected) 
planet distances from the main star of
$ 0.96 AU  \le x_{pl} \sqrt{M_*/0.3 M_{\odot}} \le 3.8 AU$, 
very  interesting indeed for expected planets.

The strip containing the caustics  produced by the planets 
in Figure 2  is displayed in Figure 3 
in higher resolution and  subtracted by 
the magnification that would
be produced by the ``main" star alone.
All the maps here and below are such ``difference" magnification maps. 
A grey level darker than average indicates magnification relative to
the case of an isolated star, brighter than average means
demagnification relative to the case of a           star without planet.
Figure 3a indicates the case of $m_{pl} = 10^{-4} M_*$, 
Figure 3b that of $m_{pl} = 10^{-5} M_*$. 
Here the light deflection  of all the eight planets is considered
at the same time. 
The horizontal lines  
mark the tracks along which light curves are determined.

In Figure 4 light curves are displayed that correspond to
sources with radius  $\theta_S = 1 * 10^{-3} \theta_E$
and which  move (relative motion compared to the planet star system!) 
along the tracks marked 
with lines in Figure 3.
For comparison, the solid line indicates the ``unperturbed" lightcurve
by an isolated star without planets.
Note that these lightcurves
are determined for all the eight planets simultaneously
being at the given positions.
It can be seen that the lightcurves are affected by more than one 
planet.
(All other lightcurves displayed below are calculated for a single
planet only.)
The time scale is in units of $t_E$. 
Note that the 
features produced by the planets can have quite high
amplitudes, their widths are  a few percent of an Einstein
radius, i.e. they can last from a few hours up to about a day or
so, for the typical values of velocity and
distances used      in Chapter 2.

\subsection*{4.2 Quantitative Features: Inside and Outside the Einstein Ring}

Subsequently we show in more detail and more quantitatively 
magnification patterns and light curves. 
In Figure 5 the magnification
effects for planets at distances of
$x_{pl}/R_E \approx 0.57, 0.86,  1.16 $ and 1.77 (cf. Figure 1b),
resulting in
positions of caustics in source plane: $\theta/\theta_E =  -1.2$, 
-0.3, 0.3, 1.2, are shown  separately, from left to right.
The three rows show the magnification patterns for three mass values: 
$m_{pl}/ M_* = 10^{-5}, 10^{-4}, 10^{-3}$ (from top to bottom).

Superimposed are contour lines that indicate differences in 
magnification relative to the case without a planet. The 
dashed line  is the ``average" magnification, dotted
lines indicate demagnification relative to the case with no planet
by $\Delta m = 0.05$ mag and 0.30 mag, solid 
contours indicate magnifications of 
$\Delta m  = $-0.05 mag, -0.30 mag, -0.50 mag.
Note the asymmetry of the contours.

Figure 6 shows the magnification patterns for these cases without
the contours, but with 4 straight lines indicating the tracks along
which lightcurves are determined and displayed in Figure 7. The three 
rows reflect the different mass 
ratios ($m_{pl}/M_* = 10^{-5}, 10^{-4}, 10^{-3}$ from top to bottom).
The four columns correspond to the four values of planet-star distances
mentioned above.
We tried to avoid tracks parallel or perpendicular to the line
planet-star, and also not to cross the caustic pattern ``centrally", in
order not to get an artificially large number of symmetric or
high amplitude light curves.

In Figures 7 various ``realistic" light curves are displayed
for circular sources with radii $\theta_S = 1 * 10^{-3} \,  \theta$
(again, the smooth, bell-shaped light curve which would
be produced by the lensing star alone is subtracted, shown
is the ``difference light curve"). 
In Figure 7a examples for 
planet masses of $m_{pl}/M_* = 10^{-5}$ are shown. 
The four columns reflect the four different planet distances, 
and the four rows are the tracks 1 to 4. 
In Figure 7b (and 7c) the same is shown for 
$m_{pl}/M_* = 10^{-4}$ (and $m_{pl}/M_* = 10^{-3}$). What is very
obvious is the large variety of light curves, shapes, durations, 
structures. The deviations range from very smooth to very peaky, from
single bumps to multiple ups and downs. Nice symmetric, double 
peaked, M-shaped events are visible as well as highly asymmetric
examples.

\subsection*{4.3 ``Resonant Lensing": Planets Close to the Einstein Ring}

Here we will explore in some detail the case of ``resonant" lensing.
This is a situation in which the planet is at or very close
to the Einstein ring of the lensing star. 
Such a situation corresponds to the case of an isolated single
lens with an external shear of $\gamma = 1.0$, a diverging
case that is not easy to treat analytically.
We place lenses at distances of 
$x_{pl}/R_E = 0.905,  0.951, 0.975, 1.025, 1.051, 1.105 $
as illustrated in Figure 1c.

The resulting caustic patterns  in the source plane are
located at normalized distances 
of -0.20, -0.10, -0.05, +0.05, +0.10, +0.20.
These non-linear combinations of the  caustics of star  plus planet 
appear quite bizarre and can create very unusually looking,
often multi-peaked light curves. 
The amplitudes, though, are quite
modest, so that resonant lensing is certainly not the most efficient
way of detecting planets with lensing. 
In Figure 8a the magnification
structures including the caustics are shown for the six cases
mentioned above with $m_{pl} = 10^{-4} M_{\odot}$ (same in Fig.8b 
for $m_{pl} = 10^{-3} M_{\odot}$). 
As in Figures 5, we display
``difference magnification maps" in the source plane, 
the magnification of the case star plus planet minus the 
magnification of the star alone. The coordinates are in Einstein radii
in the source plane.  
The darker the color, the higher the magnification relative to
the situation without planet. Note also the very bright and almost
white regions,  which indicate de-magnification relative to the
case with no planets.

In Figure 9  there are lines superimposed on the magnification
patterns for the case of ``resonant lensing", indicating the tracks
for which light  curves are determined:  left/right columns
for $m_{pl}/M_* = 10^{-4}/10^{-5}$. The tracks were chosen not
to be parallel to one of the axes, and not to purposely pass through
a cusp point, which might result in a not representative
very high magnification.
The six rows represent the six different
values for the planet positions (decreasing from top to bottom).

The light curves along these tracks are displayed in Figure 10.
The four columns represent the four different tracks, the
six rows reflect the different star-planet distances, as
indicated.
The case 
Figure 10a (10b) shows the light curves
for the mass ratio $m_{pl}/M_* = 10^{-4} (10^{-3}) $, 
for circular sources with radii $\theta_S = 10^{-3} \,  \theta_E$.

The most remarkable thing about these light curves is the wealth of
their structures. Ranging from little dips, small smooth peaks,
with or without depression before the increase, to symmetric
double peaks with depressed interior region, through asymmetric
decrease-increase features 
to
very complex multi-peak, multi-depression  shapes, sometimes looking
like two clearly separated  structures.
The duration of these noticable perturbations of the lightcurves
of the lensing stars
ranges from about one or two percent
of the time unit $t_E$ to a few dramatic cases which affect the lightcurve
for a fraction of more 
than 15\% $t_E$.
This translates into durations from shorter than one day, to about
a week at most, for the numbers used  in Chapter 2.

%
%
%
%
\section{Summary and Conclusion}

Gravitational microlensing in the Galaxy provides the opportunity 
to detect planetary companions of stars. Planets of 
Saturn mass (or higher) are comparably easily detectable, 
Uranus-type planets  are certainly possible to detect as well, and
even Earth-mass planets can be deteced in this way,  if the photometry
is accurate enough and densely enough sampled. 
We present here magnification maps produced by star plus
planet systems for a variety of planet-star distances and
planet-star mass ratios. 
These two-dimensional distributions show a wealth of 
caustic patterns of the planet-star system. 
We also present a variety of example light curves for finite sources,
along tracks with various impact parameters and directions for all these
cases.
In particular we explore the case of ``resonant lensing", where
the planet is located very close to the Einstein ring radius of the
main star. Such a configuration gives rise to particular 
interesting caustics (cf. Figs. 8a/b)

For a source radius of $\theta_S =  10^{-3}\, \theta_E$,
(roughly a solar radius at a distance of a lensed star at 8kpc),
we determine light curves, which show a very large range of complexity.
The effect of planet lensing can produce peaks and dips in the
smooth light curve of a single star alone, i.e. local magnification
and de-magnification, relative to the case without  planet.
The perturbation/modulation  of the smooth lightcurve by
the planet can cause a large variety of
structures. Amplitudes of a few percent for earth type planets  and
higher for more massive ones  can be found for a fair fraction
of the light curves that pass near the location of the caustic.
The caustics of the planet fall inside the Einstein ring of the main 
star for star-planet separations of  
$0.618 \le x_{pl}/R_E \le 1.618$, which
translates into physical distances of 
$1.19 AU \le x_{pl} (M_*/0.3 M_{\odot})^{0.5}  \le  3.2 AU$,
for lens (source) distances of 6kpc (8kpc). This is
quite an interesting range for planet searches.

The duration of the planet-induced variations in the
lightcurve of microlensed stars is often short, typically a
few percent of the normalized time. In physical units
this can range from about 10 hours to a few days at most. 
The current microlensing experiments (cf. introduction) do or try
online data reduction so that events can be caught in real time and
alarms can be triggered.
There are at least two groups involved
in immediate follow up programs of such alarms: 
PLANET collaboration -- Probing Lensing Anomalies NETwork  --
(see Albrow et al. 1996), 
and GMAN  -- Global Microlensing Alert Network --
(see Pratt et al. 1996).
Early warning/alert  systems (Udalski et al. 1994b; Pratt et al. 1996)
and subsequent frequent and accurate photometry   -- all of
which is in principle in place already -- are the only requirements
to make sure that sooner or later galactic
planets down to earth mass  will be found.
It is imaginable that a ``second alarm" can be issued, once the 
measured light curve starts to deviate from the predicted
theoretical light curve, which carries the high probability of 
the detection of a small-mass companion.
There are already  plans to implement such planet searches
by  on a big scale (Tytler 1995, as cited by Witt \& Mao 1996).

It appears that     a search for earth mass (or more massive)
planets by  gravitational microlensing is a viable and certainly 
comparatively inexpensive way to discover planetary  systems around
other stars.. 
In principle it can be done with the
current equipment, for higher efficiency it certainly helps
if the groups involved get larger telescopes, which allow shorter and
more frequent exposures, larger fields, which allows to monitor more
stars, and more effective on-line data reduction and
faster alarming, which increases the chance of good coverage of the
candidate light curve.

Detecting planets by microlensing, which is explored here, 
can  certainly not replace the other methods to detect
planets, e.g. the ``Doppler wobble" already successfully
employed to find planets by  Mayor \& Queloz (1995),
Marcy \& Butler (1996) or Butler \& Marcy (1996).
It is a completely independent method, though, 
which will complement the methods used already.
Searching for planets by microlensing  has some strengths: 
it will be able to find star-planet-systems quite far away 
(up to many kiloparsecs), 
and probably can extend the mass range
downward  from Jupiter-type planets 
already with current technology. 
Detecting planets by microlensing
has biases different from those of the other methods: 
Whereas the Doppler wobble technique
is most sensitive to planets circling the main star 
very closely, and the astrometric attempts
to find planets work best for planets very far out, 
microlensing can best detect planets 
in  a relatively narrow (projected) distance 
range, corresponding to a few AU for typical lens-source and 
planet-star distances. 
This may possibly be of interest
if one searches particularly for planets in a similar environment
as the earth.

Once successfully found in large enough numbers, 
planets discovered by microlensing can  be 
used extensively for statistical investigations of the kind: 
what fraction of stars with mass X has
planets with mass Y  or similarly.
Such studies  may ultimately help us understand
the formation and evolution of our solar planetary system.
A lot needs to be done theoretically, in order to provide the 
underlying groundwork for such tasks. A study naturally 
extending this one is in progress, in
determining cross sections for detecting
planets of different masses, at different distances from the star, 
and for sources of different sizes.

\section*{Acknowledgments}
It is a pleasure to thank  Bohdan Paczy\'nski 
for many valuable discussions at various stages of this project,
and Peter Friedrich for his helpful advice on the production
of the figures.
I also like to gratefully acknowledge Kim Griest, Shude Mao,
Stanton Peale, Jean Schneider and Hans-J\"org Witt
for their careful reading of the manuscript and for
providing me  with many useful comments.

\section*{Figure Captions} 
%
%
\begin{figure*}
\caption{Geometry of the various star-planet configurations that are
		investigated below. The star symbol at the origin of
		the coordinate system marks the position
		of the ``main" lensing star. The crosses indicate the
		positions of the planetary companions.
		The scale is in units of Einstein radii of the
		main lensing star. The dashed line shows   
		the Einstein ring radius.
	a) Eight planets at distances between 0.57 and 1.77 $R_E$
		(investigated more closely in Figures 2, 3, 4 );
	b) Four planets at positions 
               $ 0.57 R_E, 0.86 R_E,  1.16 R_E $ and $1.77 R_E$ 
		(explored individually in Figures 5, 6,  and 7);
	c) Six planet positions close to the Einstein ring (``resonant
		lensing"): 
		$x_{pl}/R_E \approx 0.905,  0.951, 0.975, 1.025, 1.051, 
		1.105 $ (cf. Figures 8, 9 and 10).
        }
\end{figure*}

%
%
\begin{figure*}
\caption{Magnification distribution in the source plane for a star and 
	(simultaneously) eight aligned planets at various distances 
	to the star. 
	The field ranges from -1.3 to +1.3 left to right and bottom
	to top (in units of Einstein radii). The positions
	of the planets along the x-axis are indicated in Figure 1a.
	The corresponding caustic features are located along the
	x-axis at positions
	$\theta_{caus}/\theta_{E} = -1.2, -0.9, -0.6, -0.3, +0.3, 
	+0.6, +0.9, +1.2$ in the source plane
	(Note that the
	caustics for the planets which are inside the Einstein radius,
	but on the ``right side" of the star, are projected 
	to the ``left side"  in the source plane,
	due to the light deflection by the star).
	Magnification is displayed in gray color, darker gray means 
	higher magnification. 
	The gravitational lens action of the planets produces the 
	features along the x-axis, 
	four left of the position of the star 
	(from the planets with their positions inside
	the Einstein ring) and four on the right  hand side 
	(from planets outside). 
	Note that the planetary perturbations cause some regions to be
	magnified relative to the case without planets
	(darker than environment) as well as other regions to 
	be demagnified (brighter).
	The black circle indicates the Einstein ring for the star alone.
	The left panel is for planet masses of $m_{pl} = 10^{-3} M_*$, 
	i.e. an Saturn-type mass;
	the right panel is for $m_{pl} = 10^{-4} M_*$, rather a
	Uranus-like mass for a typical $M_* \approx 0.3 M_{\odot}$.  }
\end{figure*}

%
%
\begin{figure*}
\caption{Caustic regions from Figure 2 in higher resolution.
	What is shown here is a ``difference magnification map" for
	a lens consisting  of star plus eight planets:
	the magnification for lensing by an isolated star is subtracted,
	in order to get a higher dynamic range in magnification.
	Regions that are darker than the ``average" gray 
	           are magnified relative to the case with no planet, 
	those that are brighter are demagnified.
	a) The two panels correspond to the regions left and right of 
	the origin for planet masses of $m_{pl} = 10^{-4} M_*$, 
	respectively. 
	b)                          same for $m_{pl} = 10^{-3} M_*$.
	Note the bright regions between the triangular caustics 
	and directly outside the fold caustics: they represent
	positions that are demagnified relative to the case 
	of lensing by a star alone.
	The thin black lines indicate the tracks for which 
	light curves are displayed in Figure 4.
	}
\end{figure*}

%
%
\begin{figure*}
\caption {Lightcurves for the tracks indicated in Figures 3.
        a) planet masses $m_{pl} = 10^{-4} M_*$); 
	b) planet masses $m_{pl} = 10^{-3} M_*$.
	The solid line is the light curve that would be obtained 
	without a planetary companion. 
	The dashed curves indicate the ``top" track in Fig.3, the
	dotted curves indicate the ``bottom" track. 
	(Note that the left and right wings of the light
	curves may arise from different tracks.)
	The source is assumed to be a  homogeneous disk
	with radius $\theta_S \approx 1*10^{-3} \theta_E$.
	The influence of planets on the lightcurve
	of a microlensed star can be upward and downward 
	changes relative to the undisturbed case.
	}
\end{figure*}

%
%
\begin{figure*}
\caption {High resolution magnification patterns for various
	values of planet mass $m_{pl}$ and planet-star distance 
	$x_{pl}$. 
	Shown is the magnification of the star-planet-system,
	minus the magnification of the star alone. 
	The three rows
	correspond from top to bottom to planet masses of 
	$m_{pl} / M_* = 10^{-5} $, $10^{-4}$, 
	and $10^{-3} $.
	The four columns represent  from left to right
	planet distances of 
        $ x_{pl} \approx 0.57 R_E, 0.86 R_E,  1.16 R_E $ and 
	$1.77 R_E$ in the lens plane 
	(cf. Figure 1b). 
	This corresponds to positions of the centers of the caustics 
	(in the source plane) of  -1.2, -0.3, 0.3, 1.2. 
	On top of the qualitative gray scale 
	(darker means higher magnification), there
	are contour lines for more quantitative assessment: The
	dashed line means $\Delta m  = 0.0 mag$,
	the solid lines indicate values higher than that of
	an isolated   star: $\Delta m/mag =  -0.05, -0.30, -0.50$,
	dotted contours mean demagnification
	relative to this case
	$\Delta m/mag =  0.05, 0.30$. 
	}
\end{figure*}

%
%
\begin{figure*}
\caption {The same magnification patterns as shown in Figure 5, 
	here with four straight lines indicating four different tracks 
	along which light curves are determined.
	Rows from top to bottom correspond to planet masses 
	$m_{pl}/M_*= 10^{-5}, 10^{-4}, 10^{-3}$, respectively. 
	Columns from left to right represent planet-star distances 
	$x_{pl}/R_E  \approx 0.57, 0.86, 1.16 $ and $1.77$.
	All panels are the same size, 0.3 $R_E$ at a side, centered
	at the positions of the caustics.
	The corresponding light curves are shown in Figure 7.
	}
\end{figure*}

%
%
\begin{figure*}
\caption {``Difference" light curves for circular sources with
	radii $\theta_S = 10^{-3} \theta_E$ along various tracks: 
	a) For planet masses $m_{pl}/M_* = 10^{-5}$ 
		(corresponding to the top row in Figure 6);
	The four columns reflect    different planet-lens distances:
	$x_{pl}/R_E = 0.57, 0.86, 1.16 $ and $1.77$
	(from left to right).
	The four rows correspond to the four tracks indicated in
	the respective  magnification patterns in Figure 6.
	b) Same for $m_{pl}/M_* = 10^{-4}$,
		cf. middle row in Figure 6;
	c) Same for $m_{pl}/M_* = 10^{-3}$,
		cf.  bottom row in Figure 6.
	}
\end{figure*}

%
%
\begin{figure*}
\caption {Magnification patterns for ``resonant lensing",
	a star with a planetary companion of mass 
	$m_{pl} = 10^{-4} M_*$ 
	near the star's Einstein ring.
	Shown are six configurations for 
	the lens positions indicated in Figure 1c. 
 	Displayed is the ``difference magnification", i.e.  
	the magnification of the system ``star plus planet" minus 
	the magnification of an ``isolated star".
	Dark (bright) gray means magnification higher (lower) than with 
	star alone.
	Coordinates are in Einstein radii of the star, with its
	position at the origin.
	b) Same for $m_{pl}/M_* = 10^{-3}$.
	}
\end{figure*}
%
%
\begin{figure*}
\caption {Magnification patterns as shown in Figure 8, 
	          with four straight lines 
	indicating four different tracks along
	which light curves are determined.
	Rows  from top to bottom: planet-star distance 
	$x_{pl}/R_E  \approx
	0.905, 0.951, 0.975, 1.025, 1.051, 1.105 $(cf. Figure 1c).
	Left column: planet mass $m_{pl}/M_* = 10^{-4}$.
	Right column: planet mass $m_{pl}/M_* = 10^{-3}$.
	All panels are the same size, 0.3 $R_E$  by  0.15 $R_E$ 
	centered at the positions of the caustics.
	The corresponding light curves are shown in Figure 10.
	}
\end{figure*}

%
%
\begin{figure*}
\caption{a) Light curves for planet mass $m_{pl} = 10^{-4} M_*$ and
	planet positions near the Einstein ring 
	radius (cf. Figures 1c), along the tracks indicated in Figure 9.
	The source sizes are $\theta_S = 10^{-3} \theta_E$.
	The six rows correspond to the six magnification patterns
	in Figure 9 with increasing distance of the planet 
	(between $x_{pl}/R_E = 0.9 $ and 1.1).
	The four columns represent the four tracks.
	b) Same for planet mass $m_{pl}/M_* = 10^{-3}$.
	}
\end{figure*}

\newpage
\bsp

\label{lastpage}

\end{document}